\documentclass[preprint,showpacs,aps]{revtex4}
\usepackage{dcolumn}
\usepackage{graphicx}
\begin{document}

\title{Enhanced Radiative Transition in Si${}_{n}$Ge${}_{m}$ Nanoclusters} 

\author{Ming Yu, C. S. Jayanthi, David A. Drabold${}^{\ddagger}$, and S. Y. Wu}
\address{Department of Physics, University of Louisville,
Louisville, KY, 40292\\
${}^{\ddagger}$ Department of Physics and Astronomy, 
Condensed Matter and Surface Sciences Program,
Ohio University, Athens, OH 45701-2979}

\date{\today}

\begin{abstract}
Using an {\it ab-initio} molecular dynamics scheme (the Fireball scheme), 
we determined the equilibrium 
structure of intermediate size Si${}_{n}$Ge${}_{m}$ ($n+m=71$) 
nanoclusters with/without 
hydrogen passivation on the surface. Due to the strong surface 
distortion, defect states are found to permeate the energy gap of 
Si${}_{n}$Ge${}_{m}$ clusters. However, the defect states are removed 
by adding H atoms on the surface of Si${}_{n}$Ge${}_{m}$ clusters, 
and the gap opens up to a few eV, indicating a blueshift for 
photoluminescence. It is also found that the radiative transition between the 
highest occupied molecular orbital (HOMO) and the lowest unoccupied molecular 
orbital (LUMO) states is enhanced by one to two orders of magnitude for 
Si${}_{n}$Ge${}_{m}$ nanoclusters 
with respect to the corresponding pure Si clusters. This significant 
increase of the emission probability is attributed to the strong overlap 
of HOMO and LUMO wavefunctions that are centered mostly on the Ge atoms.                
\end{abstract}


\maketitle

\newpage
\section{Introduction}  
The optical properties of bulk Si and Ge are rather mediocre because 
the light emission in the bulk Si and Ge is a phonon-assisted
indirect process. Therefore, to improve on the light emission feature 
in Si-based materials is a challenge for both the technological
and the fundamental research. 

The luminescence is a result of a significant overlap in electron and hole
wave functions as the strength of the luminescence ({\it i.e.}
the emission rate and
quantum efficiency) depends on the extent of this overlap and the transition
probability. Possible means to increase this overlap 
for the Si-based materials may be accomplished through, for example, 
alloying to change the band structure, impurities to produce the intermediate
state through which the electron can recombine with the hole, or zone folding
to yield the desired quasi-direct transition \cite{iyer1}. However, the
most important breakthrough in this issue is the observation of the visible 
photoluminescence (PL) from porous Si \cite{canham,sagnes,colcott} 
and Si quantum dots \cite{takagi} which opens new possibility 
for fabricating visible 
light emitting device from Si-based materials. Structural analysis of 
porous Si is quite difficult. But several measurements have
confirmed that the principal feature of porous Si consists of extremely fine
structures which are small enough to exhibit quantum confinement effects 
\cite{canham,colcott,takagi}. 
Various theoretical works have been reported on Si nanowires 
\cite{allan,sanders,buda,hill,proot} and Si clusters 
\cite{delley,chelikowsky1,chelikowsky2,chelikowsky3,angel}. They 
clarified that the quantum confinement effects give rise to the
change in the electronic structure and the optical properties and are
the principle mechanism of the blueshift PL in porous Si and Si
quantum dots. 

There were experimental reports indicating that Ge quantum dots
embedded in SiO${}_{2}$ glassy matrices \cite{yoshihiko} 
or in porous Si \cite{feng} show a strong room temperature luminescence. 
Theoretical studies on the structure and stability of Ge clusters 
\cite{pizzagalli}, the polarizabilities of small Ge clusters 
\cite{chelikowsky4}, and the quantum confinement effect on excitons 
in Ge quantum dots \cite{toshihide} have also been reported. 
Since Ge has smaller electron and hole effective masses and a
larger dielectric constant than the corresponding quantities for
Si, the effective Bohr radius of
the exciton in Ge is larger than that in Si, and the quantum confinement
effect appears more pronounced in Ge than in Si 
\cite{feng,yoshihiko,toshihide,yosuke}. 
These results suggest that Si${}_{n}$Ge${}_{m}$ nanoclusters could be 
possible candidates to be used as components for nanoscale 
functional optical devices. 
In order to understand the physics of Si${}_{n}$Ge${}_{m}$
clusters, we performed an {\it ab initio} molecular dynamics 
simulation for Si${}_{n}$Ge${}_{m}$ clusters of an intermediate size, 
and systematically studied their electronic
and optical properties. There is no doubt that the mismatch effect
dominating the electronic and optical properties in
Si${}_{1-x}$Ge${}_{x}$ alloys may introduce
interesting optical features in Si${}_{n}$Ge${}_{m}$ clusters. 
But one has to keep in mind that
the surface distortion associated with the stabilizing of the 
Si${}_{n}$Ge${}_{m}$ clusters will also play an important role. 
Therefore the competition between the lattice  
mismatch and the surface distortion is the basic issue in our investigation.
For this purpose, we studied Si${}_{n}$Ge${}_{m}$ clusters of
an intermediate size with/without the hydrogen passivation. 
By comparing our results between the two sets of Si${}_{n}$Ge${}_{m}$ clusters, 
we found that the surface distortion plays an important role in the 
intermediate size of Si${}_{n}$Ge${}_{m}$ clusters 
but the mismatch effect dominates when  
Si${}_{n}$Ge${}_{m}$ clusters of the intermediate size are passivated by 
hydrogen atoms to eliminate the dangling bonds so as 
to lessen the effect of the surface distortion. We also found that
the latter shows an enhancement of radiative
transition and a blueshift in PL.

It should be noted that our simulations of the Si${}_{n}$Ge${}_{m}$ clusters without
hydrogen passivation will lead only to one of the more stable
configurations among a large number of structural isomers. Hence
its resulting structural and electronic properties may not exactly
represent the corresponding properties of the true ground state
configuration of the Si${}_{n}$Ge${}_{m}$ clusters. However this 
caveat does not affect
the main conclusions of our study, including the role played by the lattice
mismatch and the surface relaxation in Si${}_{n}$Ge${}_{m}$ 
clusters with/without hydrogen
passivation, and the enhanced radiative 
transition in Si${}_{n}$Ge${}_{m}$ clusters.
 
\section{Method} 

We considered 71-atom Si${}_{n}$Ge${}_{m}$ clusters, with $m$ taking on 
the values 0, 18, 35, 53, and 71, respectively. The initial configurations 
for the five clusters were generated randomly on a regular tetrahedral
network. They were then relaxed by the {\it ab initio} molecular
dynamics scheme developed by Sankey and co-workers (the Fireball scheme) 
\cite{sankey}.
This scheme is based on the density functional theory (DFT) in the
local density approximation (LDA), where a local basis set is
used to construct the Kohn-Sham orbitals. The basis functions
are slightly excited pseudo-atomic orbitals (PAO). The Kohn-Sham
orbitals are calculated self-consistently using the
Hamann-Schl\"uter-Chiang pseudo-potential \cite{hamann} and
the Ceperley-Alder form of the exchange-correlation potential
as parameterized by Perdew and Zunger \cite{perdew}. We have tested this
method in the study of the strain relaxation of Si${}_{1-x}$Ge${}_{x}$
alloys and the results are in good agreement with the experimental
observations \cite{yu2}. \\

In our simulation, $sp^{3}$-type PAOs were used with confinement
radii of 5.0 $a_{\rm B}$, 5.2 $a_{\rm B}$, and 3.6 $a_{\rm B}$
for Si, Ge, and H atoms, respectively. A cubic cell with a lattice 
constant 50 \AA $\;$ was chosen as the unit cell. This cell size 
is sufficiently large to ensure that spurious interactions
between clusters will vanish. To obtain the stable configuration
of a Si${}_{n}$Ge${}_{m}$ or a Si${}_{n}$Ge${}_{m}$H${}_{84}$
cluster, molecular dynamics simulations were performed on the clusters, 
starting from the initial random configuration, at a
temperature of 10${}^{3}$ K for about 20 ps until the network was equilibrated.
The network was then slowly cooled to 300 K for 5 ps, and
the dynamical quenching was finally performed to fully
relax the system to zero K. Charge transfer was
calculated self-consistently in the simulations, which is important
in modeling clusters.\\

After performing the molecular dynamical simulations, we obtained 
stable configurations, five each for Si${}_{n}$Ge${}_{m}$ and 
Si${}_{n}$Ge${}_{m}$H${}_{84}$ clusters, with
the ratio $m/(n+m)$ = 0.0, 0.25, 0.51, 0.75, and 1.0, respectively.
As an example, Fig.~\ref{fig1} (a) shows the stabilized configuration 
of the Si${}_{36}$Ge${}_{35}$ cluster. Apparently, there is strong surface 
distortion in this case and the
stabilized structure is a compact network in a more oblate structure. 
This kind of structure has also been found to be more stable for the 
intermediate-size of Si clusters from other theoretical studies \cite{ho}. 
It is found that the surface distortion leads to 
local bonding configurations with more than four bonds, in particular
in the vicinity of the surface. Further structural
analysis shows that the average bond lengths of Si-Si (b${}_{\rm SiSi}$), 
Si-Ge (b${}_{\rm SiGe}$), and Ge-Ge (b${}_{\rm GeGe}$) of 
the Si${}_{n}$Ge${}_{m}$ clusters are expanded relative to the bulk, 
as shown in Fig.~\ref{fig2} (a). But they still maintain the relationship of
$b_{\rm SiSi} < b_{\rm SiGe} < b_{\rm GeGe}$ and are almost independent
of the ratio of $m/(n+m)$. In addition, as shown in Fig.~\ref{fig2} (b), 
the average angles $\theta_{\alpha \beta \gamma}$, which shows the
angle between $\beta \alpha$ and $\beta \gamma$ atoms, 
are less than the tetrahedral angle,  
except $\theta_{\rm Si Si Si}$ in Si${}_{18}$Ge${}_{53}$ cluster.
The large fluctuation  of the average angles $\theta_{\alpha \beta \gamma}$ 
as a function of the ratio of $m/(n+m)$ indicates that the clusters have 
lost their initial tetrahedral symmetry. This means that the surface 
distortion strongly dominates in the intermediate size 
Si${}_{n}$Ge${}_{m}$ clusters, whatever the ratio of Si/Ge. 

When the initial configurations were passivated 
with 84 H atoms to terminate the dangling bonds on the surface atoms, 
all the relaxed Si${}_{n}$Ge${}_{m}$H${}_{84}$ clusters 
still maintain a tetrahedral symmetry in the interior, with only a minor 
distortion on the surface. Such features can be seen, for example, 
from the relaxed structure of Si${}_{36}$Ge${}_{35}$H${}_{84}$
shown in Fig.~\ref{fig1} (b). The average bond lengths b${}_{\rm SiSi}$, 
b${}_{\rm SiGe}$, and b${}_{\rm GeGe}$ of the cluster (see Fig.~\ref{fig2} (c))
and the average angles $\theta_{\alpha \beta \gamma}$ 
(see Fig.~\ref{fig2} (d)) are quite close to the corresponding features in 
Si${}_{1-x}$Ge${}_{x}$ alloys (comparing Figs.~\ref{fig2} (c) and (d) 
to Figs. 3 and 7 in ref. \cite{yu2}). This is an indication 
that the surface distortion in hydrogen-passivated clusters
is weak even for Si${}_{n}$Ge${}_{m}$H${}_{l}$ clusters of an intermediate size. 
On the other hand, the mismatch effect becomes the dominating factor in 
hydrogenated Si${}_{n}$Ge${}_{m}$ clusters, similar to the situation exhibited 
in Si${}_{1-x}$Ge${}_{x}$ alloys \cite{yu2}.\\

\section{Electronic Structure}

The LDA calculation is known to underestimate the energy gap in the
single-particle energy spectrum of semiconductor clusters.
Corrections to the underestimated energy gap can be calculated using
the GW approach \cite{chelikowsky3,gw,yu1}. However, GW calculations 
are expensive, even for clusters of an intermediate size considered
in this work (i.e., Si${}_{n}$Ge${}_{m}$H${}_{84}$ clusters
with $n+m=71$). Our main interest in this study is to provide
an understanding of the trend of change in the electronic
structure and optical properties of Si${}_{n}$Ge${}_{m}$
clusters of a given size as the concentration of the Ge component
varies. The LDA approach is expected to be sufficiently 
accurate to provide a qualitative description of this trend of change.
Therefore, we have calculated the electronic density of states (EDOS)
of the relaxed  Si${}_{n}$Ge${}_{m}$ clusters, using the same DFT/LDA scheme
as that in the structural determination.

The EDOSs of Si${}_{n}$Ge${}_{m}$ 
and Si${}_{n}$Ge${}_{m}$H${}_{84}$ clusters at the five ratios of $m/(n+m)$
described previously are shown in Fig.~\ref{fig3} . From the left panel of 
Fig.~\ref{fig3}, many defect states associated with the surface distortion were
seen in the HOMO-LUMO energy gap region of Si${}_{n}$Ge${}_{m}$ clusters.
These defect states were cleared up when the surface was passivated by H atoms
(see the right panel of Fig. 3) because H atoms terminate the dangling bonds 
of the surface atoms, resulting in the clusters keeping the basic tetrahedral 
symmetry. Even though the energy spectrum of Si${}_{n}$Ge${}_{m}$
and that of Si${}_{n}$Ge${}_{m}$H${}_{84}$ clusters are quite different,
the general feature of the energy spectrum in both cases was found to 
be rather insensitive to the ratio of $m/(n+m)$.

Fig.~\ref{fig4} (a) and (b) illustrate the HOMO-LUMO energy gap of 
Si${}_{n}$Ge${}_{m}$ and Si${}_{n}$Ge${}_{m}$H${}_{84}$ clusters
as a function of the ratio of $m/(n+m)$, respectively.
Because of the existence of the defect states in the energy gap region, 
the HOMO-LUMO gap of Si${}_{n}$Ge${}_{m}$ clusters is small 
(only 0.4-0.5 eV) and does not show any dependence on the ratio of 
Si to Ge. But the HOMO-LUMO gap of the hydrogenated Si${}_{n}$Ge${}_{m}$ 
clusters is opened up to several eV (3.5 - 4.0 eV) and shows a linear 
dependence on the ratio of Si to Ge which reflects the mismatch 
effect similar to that for the corresponding alloys \cite{yu2}. 
Specifically, experimental observation \cite{rubin} and
theoretical calculations \cite{shen,robert} indicate that the
lattice mismatch associated with alloying in Si${}_{1-x}$Ge${}_{x}$ alloys
induces a change of the conduction band, in particular,
the lowest energy in the conduction band (LUMO) changes from
the ${\bf \Delta}$ point near X point for Si-rich alloys to the L point
for Ge-rich alloys \cite{robert}, leading to a nearly linear decrease of
the indirect band gap. In the case of Si${}_{n}$Ge${}_{m}$ 
clusters passivated by
hydrogen, the dominating factor is again
the lattice mismatch. This similar effect is
observed from the HOMO-LUMO gap as shown in Fig.~\ref{fig4}b.
The existence of such large energy gap for Si${}_{n}$Ge${}_{m}$ clusters 
of this size (with $n+m=71$) can be attributed to the effect of 
quantum confinement and is consistent with the scenario of
a blueshift in PL.\\

\section{Optical Properties}
 It is well known that an accurate calculation of the band gap
is important for the determination of the optical polarizability of
semiconductors and insulators. Many efforts have been focused on the
correction to the "too-small" band gap obtained by DFT/LDA calculations
for the purpose of accurately predicting the optical response of
these systems. Another key factor for determining the optical properties
of these systems is a proper description of interacting electron-hole
pairs (excitons). While calculations using the Green's function
based on the GW approach have led to quite accurate prediction of
the energy of low-lying excitations and the oscillator strength
for small clusters \cite{rohlfing}, 
the application of this method to clusters of
intermediate sizes is still computationally too expensive.
Most recently, a method based on the linear-response theory within the
frame work of the time-dependent (TD) DFT/LDA had been applied
to calculate the optical spectra of clusters \cite{chelikowsky2} 
with results in general agreement with experimental 
results as well as more complicated
theoretical methods (e.g., GW approach). However, at present, the
application of the method is still limited to clusters of radius less than
1 $nm$.

The study of the dielectric function via the optical transition
matrix elements allows us
to obtain information directly on absorption and photoluminescence
spectra. It can also allow us to obtain information
indirectly on the relevant radiative PL processes.
The imaginary part of the dielectric function can be calculated
by \cite{rohlfing} 

\begin{eqnarray}  
{\rm Im}\epsilon (\omega)= \frac{16 \pi e^{2}}{\omega^{2}}
\sum_{S} | \hat{\bf e} \cdot <0 | {\bf v} | S> |^{2}
\delta (\omega-\Omega_{S})
\label{eq:eq1}
\end{eqnarray} 
\noindent  
where $\hat{\bf e} $ is the polarization vector of the light, 
${\bf v}=i/\hbar [H, {\bf r}]$, $|0>$ refers to the ground state, 
$|S>$ the excited state, and $\Omega_{S}$ the excitation angular frequency. 
The excited state can be expanded
in electron-hole pair configuration such that

\begin{eqnarray}  
|S>= \sum_{v}^{\rm hole} \sum_{c}^{\rm elec} A^{S}_{vc} |vc>
\label{eq:eq2}
\end{eqnarray} 
\noindent  
The coupling coefficient $ A^{S}_{vc}$ can be calculated within the
framework of the two-particle Green's function by solving the 
corresponding Bethe-Salpeter equation. A reasonable approximation
leads to the determination of $ A^{S}_{vc} $ as the positive solutions to the
eigenvalue equation \cite{rohlfing}.

\begin{eqnarray}  
(E_{c}-E_{v}) A^{S}_{vc} + \sum_{v',c'}K^{AA}_{vc,v'c'} (\Omega_{S}) 
A^{S}_{v'c'} = \Omega_{S} A^{S}_{vc}
\label{eq:eq3}
\end{eqnarray}
\noindent  
where $E_{c}$ is the single-electron energy in the conduction band, 
$E_{v}$ the single-electron energy in the valence band, and
$K^{AA}_{vc,v'c'}$ the electron-hole interaction kernel. The most
computationally intensive part in the calculation is the evaluation of the
electron-hole interaction kernel $K^{AA}_{vc,v'c'}$. This bottleneck
is one of the main culprits that limits the application of the
GW approach to only very small clusters and that of the
TDLAD to clusters with the radius less than 1 $nm$.

Using Eq.~(\ref{eq:eq2}), the optical transition matrix elements can
be written as
\begin{eqnarray} 
<0|{\bf v}|S>=\sum_{v}^{\rm hole} \sum_{c}^{\rm elec.} A^{S}_{vc} 
<v|{\bf v}|c> 
\label{eq:eq4}
\end{eqnarray}
\noindent 
The substitution of Eq.~(\ref{eq:eq4}) into Eq.~(\ref{eq:eq1}) shows
that the imaginary part of the dielectric function depends on the
transition matrix elements of electron-hole pair configuration
$<v|{\bf v}|c>$ as well as the coupling coefficient $ A^{S}_{vc}$.
For clusters Si${}_{n}$Ge${}_{m}$ of a given size ($n+m$=constant),
the changing composition is expected to affect more on the
transition matrix elements $<v|{\bf v}|c>$ than on the coupling coefficient
$ A^{S}_{vc}$ because of the insensitivity of the EDOSs to the change
in configuration (see Eq.~(\ref{eq:eq3}) and Fig.~\ref{fig3}). Since the main interest of this work is to understand the
effect on the optical properties of Si${}_{n}$Ge${}_{m}$  clusters
of a given intermediate size ($n+m$=constant) 
by incorporating Ge atoms into the cluster,
we therefore focus our attention on the effect on $<v|{\bf v}|c>$
due to the change in the composition in Si${}_{n}$Ge${}_{m}$  clusters,
particularly on how the change in $<v|{\bf v}|c>$  affects the
radiative transition probability of the clusters as its
composition varies.

To set up a benchmark for the analysis of the trend of the the optical
features of Si${}_{n}$Ge${}_{m}$  clusters of a given size but
varying compositions within the framework considered in this work,
we have calculated the imaginary part of the dielectric function
without the electron-hole interaction. In this situation,  Eq.~(\ref{eq:eq1})
reduce to

\begin{eqnarray}  
{\rm Im}\epsilon (\omega)= \frac{16 \pi^{2} e^{2}}{V} \sum_{v}\sum_{c}
|<v|{\bf r} \cdot \hat{\bf e} |c>|^{2}  \delta(E_{c}-E_{v} -\hbar \omega)
\label{eq:eq5}
\end{eqnarray} 
\noindent
Here the sums are over all the eigenstates $|v>$ and $|c>$.
$\bf{r}$ is the position operator, and V the volume of the
cluster. Specifically, we  calculated the average 
Im$\overline{\epsilon (\omega)}$ $\bigl(= \frac{16 \pi^{2} e^{2}}{V} \sum_{v}\sum_{c}
\frac{1}{3} (|<v|x|c>|^{2}+|<v|y|c>|^{2}+|<v|z|c>|^{2} ) 
\delta(E_{c}-E_{v} -\hbar \omega)\bigr) $ since the shapes of the relaxed 
Si${}_{n}$Ge${}_{m}$ clusters are compact and suggesting a more-or-less
isotropic behavior.\\  

Fig.~\ref{fig5}  presents the calculated Im$\overline{\epsilon (\omega)}$
at various ratios of Si to Ge in the cases with and without
H-passivation. It is found that the optical gaps, 
defined by the onset energy of the spectra edge in Si${}_{n}$Ge${}_{m}$ 
clusters, are in the range of 0.4-0.5 eV 
and are almost independent of the ratio of Si to Ge. The optical spectral
peaks are smooth and broad. The spectra display
low-energy transitions and show a tail near the optical edge. 
The tail corresponds to the defect states due to the surface 
distortion as shown in EDOS of
Si${}_{n}$Ge${}_{m}$ clusters (Fig.~\ref{fig3}). Therefore, such optical 
properties are not suitable for the application as optical devices.
The optical gaps of H-terminated Si${}_{n}$Ge${}_{m}$ 
clusters, however, are in the range of 3.3-4.1 eV 
and have a linear dependence on the
ratio of Si to Ge.  Unlike optical spectra of  Si${}_{n}$Ge${}_{m}$ 
clusters, the spectral peaks of hydrogenated  Si${}_{n}$Ge${}_{m}$ clusters 
are sharp with no tail at the edge of the optical spectra.
Such peaks near the edge reflect the high structural symmetry 
(the tetrahedral symmetry in the present study) 
of the Si${}_{n}$Ge${}_{m}$H${}_{84}$ clusters.
The large optical gaps in Si${}_{n}$Ge${}_{m}$H${}_{84}$ clusters
indicate the possibility of optical applications.
                                  
Since the luminescence is a result of significant overlap in electron and hole
wave functions, and the strength of the luminescence 
depends on the extent of this overlap and the transition probability,
we investigated the radiative transition probabilities of
Si${}_{n}$Ge${}_{m}$ and Si${}_{n}$Ge${}_{m}$H${}_{84}$ clusters. 
The state-to-state spontaneous transition probability
($W_{ij}$) (the inverse of the radiative lifetime
$\tau_{ij}$) of the first-order radiative process 
between states $|i>$ and $|j>$ is defined from the Fermi golden
rule and is given by \cite{hirao}

\begin{eqnarray}  
W_{ij}= \frac{1}{\tau_{ij}}=\frac{4 e^{2} \omega_{ij}^{3} n}
{3 h c^{3}} |<i|{\bf r}  \cdot \hat{\bf e} |j>|^{2}
\label{eq:eq6}
\end{eqnarray} 
\noindent
where $e$ and $m$ are the electron charge and mass, respectively.
$c$ the speed of light, $\omega_{ij}$ the energy difference 
(divided by $h$) between states $|i>$ and $|j>$, and $n$ the 
refractive index of Si${}_{n}$Ge${}_{m}$ clusters. 
From ellipsometry \cite{pickering} and optical-absorption 
\cite{sagnes} experiments 
for porous Si, it seems that the refractive index decreases with 
increasing porosity. Since there is no measurement of the refractive index 
for Si${}_{n}$Ge${}_{m}$ clusters, we choose $n$ to be 1 in the 
present calculation.

It is apparently from the formula of $W_{ij}$ that the
energy difference $\omega_{ij}^{3}$ as well as the dipole matrix elements
$<i|{\bf r}|j>$ dominate the spontaneous transition probability.
Table I lists the HOMO-LUMO 
spontaneous emission probabilities $W_{HL}$ $\bigl(=\frac{4 e^{2} \omega_{HL}^{3} n}
{3 h c^{3}} \frac{1}{3}( |<H|x|L>|^{2}+|<H|y|L>|^{2}+|<H|z|L>|^{2}) \bigr)$
of Si${}_{n}$Ge${}_{m}$ and  Si${}_{n}$Ge${}_{m}$H${}_{84}$ clusters.
It can be seen that $W_{HL}$'s for Si${}_{n}$Ge${}_{m}$ 
clusters are very small
($ \approx 10^{-5}$ in $ns^{-1}$). The corresponding 
radiative lifetime $\tau_{HL}$ are
quite large, ${\it i.e.}$ 608 $\mu s$ for Si${}_{71}$, 49 $\mu s$ for 
Si${}_{53}$Ge${}_{18}$, 71 $\mu s$ for Si${}_{36}$Ge${}_{35}$,
25 $\mu s$ for Si${}_{18}$Ge${}_{53}$, and 121 $\mu s$ for Ge${}_{71}$,
respectively. But the HOMO-LUMO spontaneous transition 
probabilities $W_{HL}$ of hydrogenated
Si${}_{n}$Ge${}_{m}$ clusters are about two to three orders of magnitude
larger than the corresponding ones of Si${}_{n}$Ge${}_{m}$ clusters. 
Their corresponding radiative lifetime $\tau_{HL}$ are therefore
shortened by about two to three orders, 
{\it i.e.}, 8.1 $\mu s$ for Si${}_{71}$H${}_{84}$, 0.12 $\mu s$ for 
Si${}_{53}$Ge${}_{18}$H${}_{84}$, 0.059 $\mu s$ for 
Si${}_{36}$Ge${}_{35}$H${}_{84}$, 0.07 $\mu s$ for
Si${}_{18}$Ge${}_{53}$H${}_{84}$, and 0.098 $\mu s$ for
Ge${}_{71}$H${}_{84}$, respectively.
The large differences in $W_{HL}$ and $\tau_{HL}$ between
Si${}_{n}$Ge${}_{m}$ and Si${}_{n}$Ge${}_{m}$H${}_{84}$ clusters
are mainly due to the energy differences of $\omega_{HL}^{3}$ between
HOMO-LUMO states in Si${}_{n}$Ge${}_{m}$ and Si${}_{n}$Ge${}_{m}$H${}_{84}$ 
clusters. As seen in Fig.~\ref{fig3}, the HOMO-LUMO energy differences
$\omega_{HL}$ of Si${}_{n}$Ge${}_{m}$H${}_{84}$ clusters are about 10 times 
larger than those of Si${}_{n}$Ge${}_{m}$ clusters.

It is noted that the radiative lifetime $\tau_{HL}$ calculated for 
Si${}_{71}$H${}_{84}$ cluster (8.1 $\mu s$) is comparable to that
of Si${}_{66}$H${}_{64}$ cluster of a similar size (6 $\mu s$)
obtained by Hirao {\it et al.} \cite{hirao}. The spontaneous transition 
probabilities $W_{HL}$ of Si${}_{71}$H${}_{84}$ cluster (0.012 $\times
10^{4} \; ms^{-1}$) is also consistent with the recombination
rate of an excited electron-hole pair in Si crystallites 
(about $10^{4} \; ms^{-1}$) for the photon 
energy of 4.0 eV at 5 K calculated by Delerue {\it et al.} \cite{allan,proot}. 
The spontaneous transition probabilities 
$W_{HL}$ is found in our calculation to be higher 
in pure Ge${}_{71}$H${}_{84}$ cluster 
($1.01 \times 10^{-2} \; ns^{-1} $) than in pure Si${}_{71}$H${}_{84}$ 
cluster ($0.012 \times 10^{-2} \; ns^{-1} $).
This result is qualitatively consistent with the results of the radiative 
decay rate of excitons in Ge quantum dots 
(about $0.2 \times 10^{-2} \; ns^{-1} $) and in Si quantum dots 
(about $0.05 \times 10^{-2} \; ns^{-1} $)
with the dot radius of 10 \AA $\;$ obtained by Takagahara {\it et al.} 
\cite{toshihide}. 

Our analysis of the radiative transition in Si${}_{n}$Ge${}_{m}$ clusters
indicates that the HOMO-LUMO spontaneous emission probability $W_{HL}$
(radiative life time $\tau_{HL}$) of Si${}_{n}$Ge${}_{m}$
and Si${}_{n}$Ge${}_{m}$H${}_{84}$ clusters is very sensitive to the
Ge content ($m/(n+m)$) in the cluster. Table I shows a sudden
and substantial increase in $W_{HL}$  once Ge atoms
are incorporated into the Si clusters. In particular for 
hydrogen-passivated Si${}_{n}$Ge${}_{m}$H${}_{84}$ clusters,
the incorporation of Ge atoms into the Si cluster can bring
about a dramatic increase in $W_{HL}$ of up to two orders of
magnitude as compared to the pure hydrogen-passivated Si cluster.
An increase in $W_{HL}$ of this magnitude for Si${}_{n}$Ge${}_{m}$H${}_{84}$
with $n+m=71$ suggests that the incorporation of Ge atoms
into hydrogen-passivated Si clusters may pave the way to dramatically
enhance the optical properties of pure Si clusters.

To shed light on the underlying physics of this dramatic change in
the HOMO-LUMO spontaneous emission probability $W_{HL}$,
we carried out a detailed analysis of factors that might affect 
$W_{HL}$. From Eq.~(\ref{eq:eq6}), it can be seen that $W_{HL}$
is controlled by the interplay between the HOMO-LUMO gap
$\omega_{HL}$ and the dipole matrix-element $<H|{\bf r}|L>$.
As shown in Fig.~\ref{fig4}, $\omega_{HL}$ for Si${}_{n}$Ge${}_{m}$
clusters of a given size does not exhibit 
any sensitive dependence on the Ge content. 
For Si${}_{n}$Ge${}_{m}$H${}_{84}$ clusters, $\omega_{HL}$ decreases
with increasing Ge content ($m/(n+m)$) and shows a linear dependence
on the Ge content. However, the range of variation for $\omega_{HL}$ 
is only by a factor of $\sim 1.22$. On the other hand, the dependence of the
dipole matrix element on $m/(n+m)$ shows a more complicated pattern
(see Table II). For example, $ \overline{|<H|{\bf r}|L>|^{2}}=
\frac{1}{3}(|<H|x|L>|^{2}+|<H|y|L>|^{2}+|<H|z|L>|^{2})$ for 
Si${}_{n}$Ge${}_{m}$H${}_{84}$ clusters shows a dramatic increase of two 
orders of magnitude once Ge atoms are incorporated into the Si cluster.
It peaks at the equal composition of Si and Ge atoms, and then reduces to
a smaller but same order of magnitude for the pure Ge cluster. Since
the range of variation for $\overline{|<H|{\bf r}|L>|^{2}}$ (over two orders
of magnitude) far surpasses that for $\omega_{HL}^{3}$ (less than one
order of magnitude), it must be the behavior of
$\overline{|<H|{\bf r}|L>|^{2}}$ that determines the pattern of behavior
for $W_{HL}$.  As shown in Table I, $W_{HL}$ for
Si${}_{n}$Ge${}_{m}$H${}_{84}$ clusters indeed 
exhibits a similar pattern as that
of  $\overline{|<H|{\bf r}|L>|^{2}}$, namely, a drastic, almost two orders of 
magnitude, increase for small Ge content, peaking at an equal composition
of Si and Ge atoms, reducing to a somewhat smaller value for the pure 
hydrogen-passivated Ge cluster. 

To shed light on the mechanism responsible for the dramatic
increase in $\overline{|<H|{\bf r}|L>|^{2}}/W_{HL}$ once Ge atoms are 
incorporated into the hydrogen-passivated Si clusters, we express the
dipole matrix element in terms of the pseudo-atomic orbitals
within the Fireball scheme \cite{sankey}, namely,

\begin{eqnarray}  
<H|{\bf r}|L>=\sum_{i \alpha,j \beta} (c^{H}_{i \alpha})^{\ast}
c^{L}_{j \beta} <i \alpha|{\bf r}|j \beta>
\label{eq:eq7}
\end{eqnarray} 
\noindent
where $c^{\lambda}_{i \alpha} $ denotes the coefficient of
expansion of the wave function $\lambda$ in the pseudo-atomic orbital
$\alpha$ at the site $i$. The matrix element $<i \alpha|{\bf r}|j \beta>$
depends only on the structural configuration of the cluster. Specifically
\begin{eqnarray} 
<i \alpha|{\bf r}|j \beta>&=&\int \phi^{\ast}_{\alpha}({\bf r}-{\bf R}_{i})
{\bf r} \phi_{\beta}({\bf r}-{\bf R}_{j})  d{\bf r} \nonumber \\
&=&\int \phi^{\ast}_{\alpha}({\bf r'})
{\bf r'} \phi_{\beta}({\bf r'}-{\bf R}_{ij})  d{\bf r'}+
{\bf R}_{i}\int \phi^{\ast}_{\alpha}({\bf r'})
\phi_{\beta}({\bf r'}-{\bf R}_{ij})  d{\bf r'}
\label{eq:eq8}
\end{eqnarray} 
\noindent
where $\phi_{\alpha}({\bf r}-{\bf R}_{i})$ is the pseudo-atomic orbital 
$\alpha$ centered at atomic site $i$, ${\bf R}_{i}$ is the position 
vector of atomic site $i$, ${\bf R}_{ij}={\bf R}_{j}-{\bf R}_{i}$, and 
${\bf r'}={\bf r}-{\bf R}_{i}$, respectively.

Since there is no substantial structural change for hydrogen-passivated
Si${}_{n}$Ge${}_{m}$H${}_{l}$ clusters of a given size ($n+m$=constant),
Eqs.~(\ref{eq:eq7}) and (\ref{eq:eq8}) then indicate that the coefficients
of expansion of the HOMO and LUMO states play the most significant role
in determining $<H|{\bf r}|L>$. In Fig.~\ref{fig6}, we plot the absolute 
value of the coefficient $c^{\lambda}_{i \alpha}$ for the pseudo-atomic 
orbital $\phi_{\alpha}({\bf r}-{\bf R}_{i})$ vs atomic site $i$ for the
HOMO and LUMO states of hydrogen-passivated Si${}_{n}$Ge${}_{m}$H${}_{84}$
($n+m=71$) clusters with various Ge content. When the patterns of behavior for
the hydrogen-passivated pure Si cluster (Si${}_{71}$Ge${}_{0}$H${}_{84}$) 
are compared with those of the hydrogen-passivated pure Ge cluster 
(Si${}_{0}$Ge${}_{71}$H${}_{84}$), it can be seen that the HOMO state
for the two clusters shows similar distribution. However,
there is a striking difference between the patterns exhibited by
the LUMO states of the two clusters. In the case of the pure Si cluster,
the expansion coefficients of the $p_{x}$-orbital give the major
contribution to the LUMO states. On the other hand, two outstanding
features differentiate the pattern of the LUMO state of the pure Ge cluster
from that of the pure Si cluster: (1) The expansion coefficients of the
$s$-orbital give the major contribution to the LUMO state. (2) The 
magnitude of the coefficients of the $s$-orbital is greater than those
of the $p_{x}$-orbital for the pure Si cluster by a factor of 2 to 3.
An examination of other figures in Fig.~\ref{fig6} shows that the
LUMO state possesses the same two features as long as the cluster
has a Ge content and these features appear mostly at the Ge sites. 
The overlap between a $s$-orbital and neighboring orbitals
is far less restrictive as compared to that between a $p_{\alpha}$
orbital with its neighboring orbitals. This fact, together with
the more pronounced strength of the coefficients for the
$s$-orbital in the LUMO state of Ge-containing clusters,
leads to a stronger overlap between the LUMO and HOMO states
of Ge-containing clusters, thus explaining the 
correlation between the appearance of these two features
and the sudden and dramatic increase in the value of
$\overline{|<H|{\bf r}|L>|^{2}}$ once Ge atoms are incorporated into the
hydrogen-passivated Si clusters. This correlation also indicates
that these two features are mainly responsible for the sudden and dramatic
increase in $W_{HL}$ once the hydrogen-passivated Si cluster possesses
a Ge content.

Finally, it should be noted that the properties of the HOMO state are
closely related to the effective mass of heavy holes while those of the
LUMO state to the average effective mass of electrons. The mass of the
heavy hole for bulk Si is reasonably close to that for bulk Ge. On the
other hand, the average effective mass of electrons for bulk Ge is a factor
of 2 smaller than that for bulk Si. It is therefore tempting to
speculate that the behavior patterns of HOMO and LUMO states are related
to the sameness of the mass of heavy hole and the difference in
the effective mass of electron for Si and Ge, respectively.
One may also add that the fact that the effective mass of Ge is two
times smaller than that of Si could be the contributing factor for
the dominant contribution of the $s$-orbital in the LUMO state for 
clusters with a Ge content.

\section{Conclusion}      

The findings of our systematic study of the structural,
electronic, and optical properties of Si${}_{n}$Ge${}_{m}$ cluster
with/without hydrogen passivation include: (1) There is only
a weak surface distortion in hydrogen-passivated clusters
Si${}_{n}$Ge${}_{m}$H${}_{l}$ of intermediate sizes.
The mismatch effect is then the dominating factor in determining
properties of those clusters. (2) The HOMO-LUMO gap for 
Si${}_{n}$Ge${}_{m}$H${}_{l}$ clusters of intermediate sizes 
opens up to several eVs, consistent with a blueshift in
PL. It shows a linear dependence on the ratio $n/(n+m)$ 
($n+m=$ constant) similar to the pattern exhibited by 
Si${}_{1-x}$Ge${}_{x}$ alloys \cite{yu2}.
(3) The HOMO-LUMO spontaneous emission probability
$W_{HL}$ of Si${}_{n}$Ge${}_{m}$H${}_{l}$ clusters is
very sensitive to the Ge content in the cluster. Once
Ge atoms are incorporated into the cluster, it can bring
about a two-order of magnitude increase in $W_{HL}$.
Our analysis attributes this dramatic
increase to the strong overlap between HOMO and LUMO states 
around Ge atoms. Our findings therefore suggest that 
hydrogen-passivated Si${}_{n}$Ge${}_{m}$ clusters may
be a viable candidate as components in optical devices.\\

\begin{acknowledgements}
This work is supported by NSF (DMR-9802274 and DMR-011284) and
DOE (DE-FG02-00ER4582).
\end{acknowledgements}

\begin{table}[here]
\caption{\label{tab:table1}HOMO-LUMO spontaneous 
emission probability $W_{HL}$ and 
radiative lifetime $\tau_{HL}$ of Si${}_{n}$Ge${}_{m}$ and
Si${}_{n}$Ge${}_{m}$H${}_{84}$ clusters}
\begin{ruledtabular}
\begin{tabular}{ccc}
System  &$ W_{HL}  $  (ns${}^{-1}$)  & $ \tau_{HL}$  (ns)\\
\hline
Without hydrogenation &  &   \\
\hline
Si${}_{71}$    &    0.164$\times 10^{-5}$  &  6.082 $\times 10^{5}$\\
Si${}_{53}$Ge${}_{18}$    &    2.049$\times 10^{-5}$  &  0.488 $\times 10^{5}$\\
Si${}_{36}$Ge${}_{35}$    &    1.415$\times 10^{-5}$  &  0.706 $\times 10^{5}$\\
Si${}_{18}$Ge${}_{53}$    &    4.078$\times 10^{-5}$  &  0.245 $\times 10^{5}$\\
Ge${}_{71}$    &    0.824$\times 10^{-5}$  &  1.214 $\times 10^{5}$\\
\hline
With hydrogenation   &  & \\
\hline
Si${}_{71}$H${}_{84}$    &    0.0123$\times 10^{-2}$  & 81.362$\times 10^{2}$\\
Si${}_{53}$Ge${}_{18}$H${}_{84}$  &  0.845$\times 10^{-2}$  & 1.183 $\times 10^{2}$\\
Si${}_{36}$Ge${}_{35}$H${}_{84}$  &  1.673$\times 10^{-2}$  & 0.598 $\times 10^{2}$\\
Si${}_{18}$Ge${}_{53}$H${}_{84}$  &  1.423$\times 10^{-2}$  & 0.703 $\times 10^{2}$\\
Ge${}_{71}$H${}_{84}$  &  1.013$\times 10^{-2}$  & 0.987 $\times 10^{2}$\\
\end{tabular}
\end{ruledtabular}
\end{table}
\begin{table}[here]
\caption{\label{tab:table2}
Dipole matrices $ \overline{|<i|{\bf r}|j>|^{2}}=\frac{1}{3}
(|<i|x|j>|^{2}+|<i|y|j>|^{2}+|<i|z|j>|^{2}) $ 
between HOMO-LUMO states (\AA${}^{2}$) of Si${}_{n}$Ge${}_{m}$
and Si${}_{n}$Ge${}_{m}$H${}_{84}$ clusters. }
\begin{ruledtabular}
\begin{tabular}{cc} 
System  & $\overline{|<H|{\bf r}|L>|^{2}} $ \qquad \\
\hline
 Without hydrogenation  &     \\
\hline
Si${}_{71}$    &  0.0329\\
Si${}_{53}$Ge${}_{18}$  & 0.322\\
Si${}_{36}$Ge${}_{35}$  & 0.238\\
Si${}_{18}$Ge${}_{53}$  & 0.791\\
Ge${}_{71}$    & 0.181\\
\hline
With hydrogenation    &\\
\hline
Si${}_{71}$H${}_{84}$   & 0.00294 \\
Si${}_{53}$Ge${}_{18}$H${}_{84}$  & 0.292\\
Si${}_{36}$Ge${}_{35}$H${}_{84}$  &0.623\\
Si${}_{18}$Ge${}_{53}$H${}_{84}$  & 0.600\\
Ge${}_{71}$H${}_{84}$   &0.464 \\
\end{tabular}
\end{ruledtabular}
\end{table}
\clearpage
\begin{figure}[here]
\caption{\label{fig1}The stabilized structure of Si${}_{36}$Ge${}_{35}$
cluster shows a compact shape (a) and that of 
Si${}_{36}$Ge${}_{35}$H${}_{84}$ cluster 
shows a spherical-like shape with the tetrahedral symmetry 
in the interior part (b).
The Si atoms are marked by the dark grey color, Ge by the black one, and
H by the light grey one.}
\end{figure}  
\begin{figure}[here]
\caption{\label{fig2}The average bond lengths of b${}_{\rm SiSi}$, 
b${}_{\rm SiGe}$, and b${}_{\rm GeGe}$ in Si${}_{n}$Ge${}_{m}$
clusters (a) and in Si${}_{n}$Ge${}_{m}$H${}_{84}$ clusters (c) 
as a function of the ratio of $m/(n+m)$. 
The average angles $\theta_{\alpha \beta \gamma}$,
the angle between two bonds $\beta \alpha$ and $\beta \gamma$
in Si${}_{n}$Ge${}_{m}$ clusters
(b) and in Si${}_{n}$Ge${}_{m}$H${}_{84}$ clusters (d) as a function of 
the ratio $m/(n+m)$ are shown in the right side of the panel.
The dotted lines in (b) and (d) are the tetrahedral angle of 109.47 degree
in the bulk Si and Ge.}
\end{figure}  
\begin{figure}[here]
\caption{\label{fig3}The electronic densities of states of Si${}_{n}$Ge${}_{m}$ clusters
(left column) and Si${}_{n}$Ge${}_{m}$H${}_{84}$ clusters (right column)
at various ratio of $ m/(n+m)$.
The Fermi levels are denoted by the vertical bars and the broaden is 0.05 eV.}
\end{figure}  
\begin{figure}[here]
\caption{\label{fig4}The HOMO-LUMO energy gap of Si${}_{n}$Ge${}_{m}$ clusters
(a) and Si${}_{n}$Ge${}_{m}$H${}_{84}$ clusters (b) as a function of
the ratio of $m/(n+m)$.}
\end{figure}  
\begin{figure}[here]
\caption{\label{fig5}The averaged imaginary part of the dielectric 
functions calculated from Eqs.~(\ref{eq:eq5}) of
Si${}_{n}$Ge${}_{m}$ clusters (left column) and 
Si${}_{n}$Ge${}_{m}$H${}_{84}$ clusters (right column)
at various ratio of $m/(n+m)$. The broaden is 0.05 eV.}
\end{figure}  
\begin{figure}[here]
\caption{\label{fig6}The absolute value of the 
coefficient $c^{\lambda}_{i \alpha}$
vs the pseudo-atomic orbital $\alpha$ centered at atomic site $i$ 
for the HOMO (left column) and LUMO states (right column) of 
Si${}_{n}$Ge${}_{m}$H${}_{84}$ clusters. 
The numbers labeled at x axis denote the number of atomic site. 
At each atomic site, there are four histograms corresponding to
the pseudo-atomic orbital $\alpha$. The order of pseudo-atomic 
orbital $\alpha$ is $s$, $p_{x}$, $p_{y}$, and 
$p_{z}$, from left to right, respectively.}
\end{figure}  

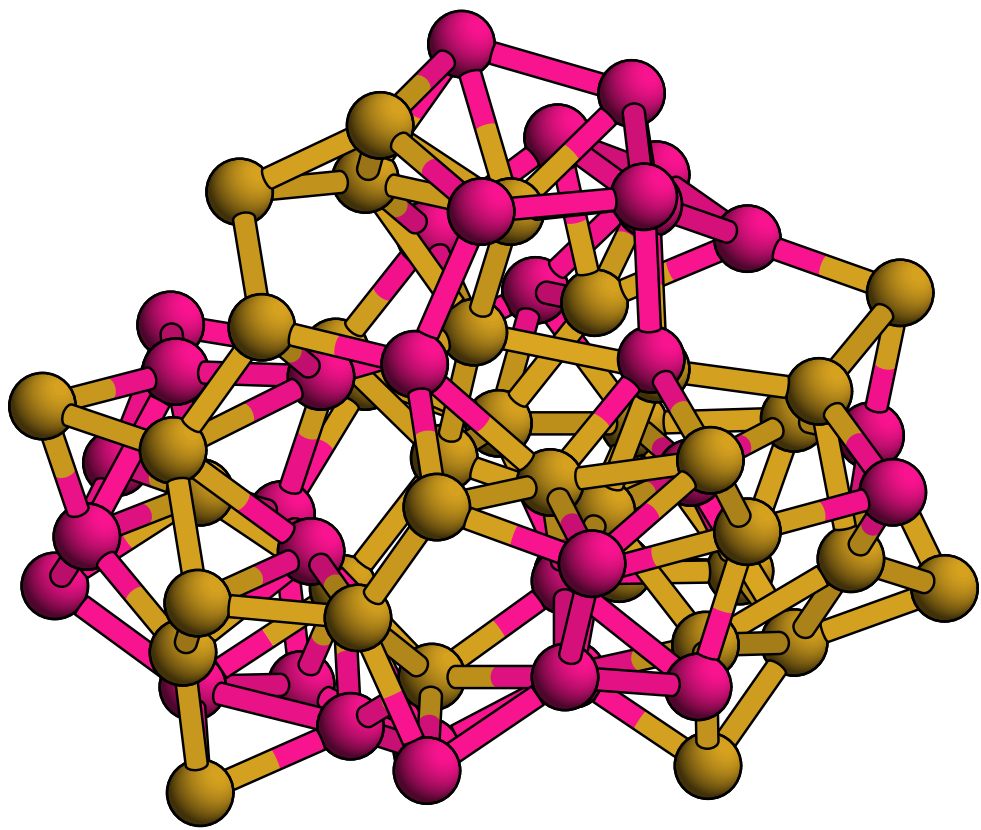
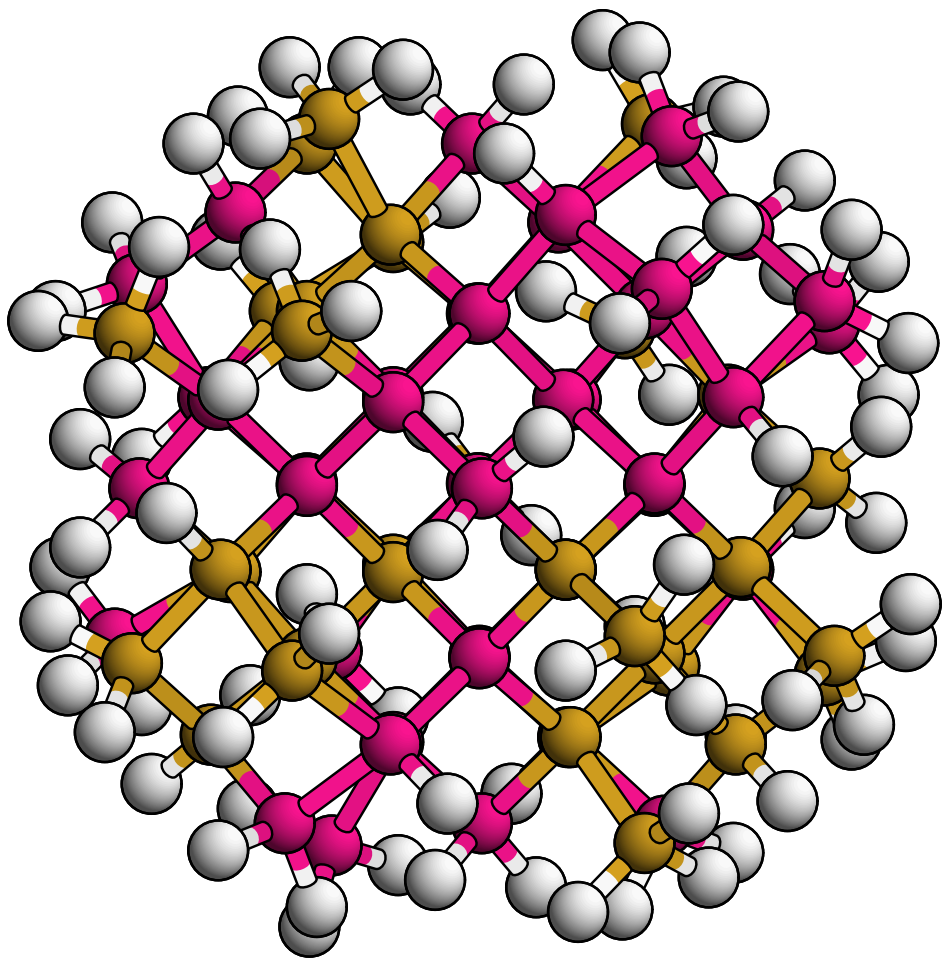
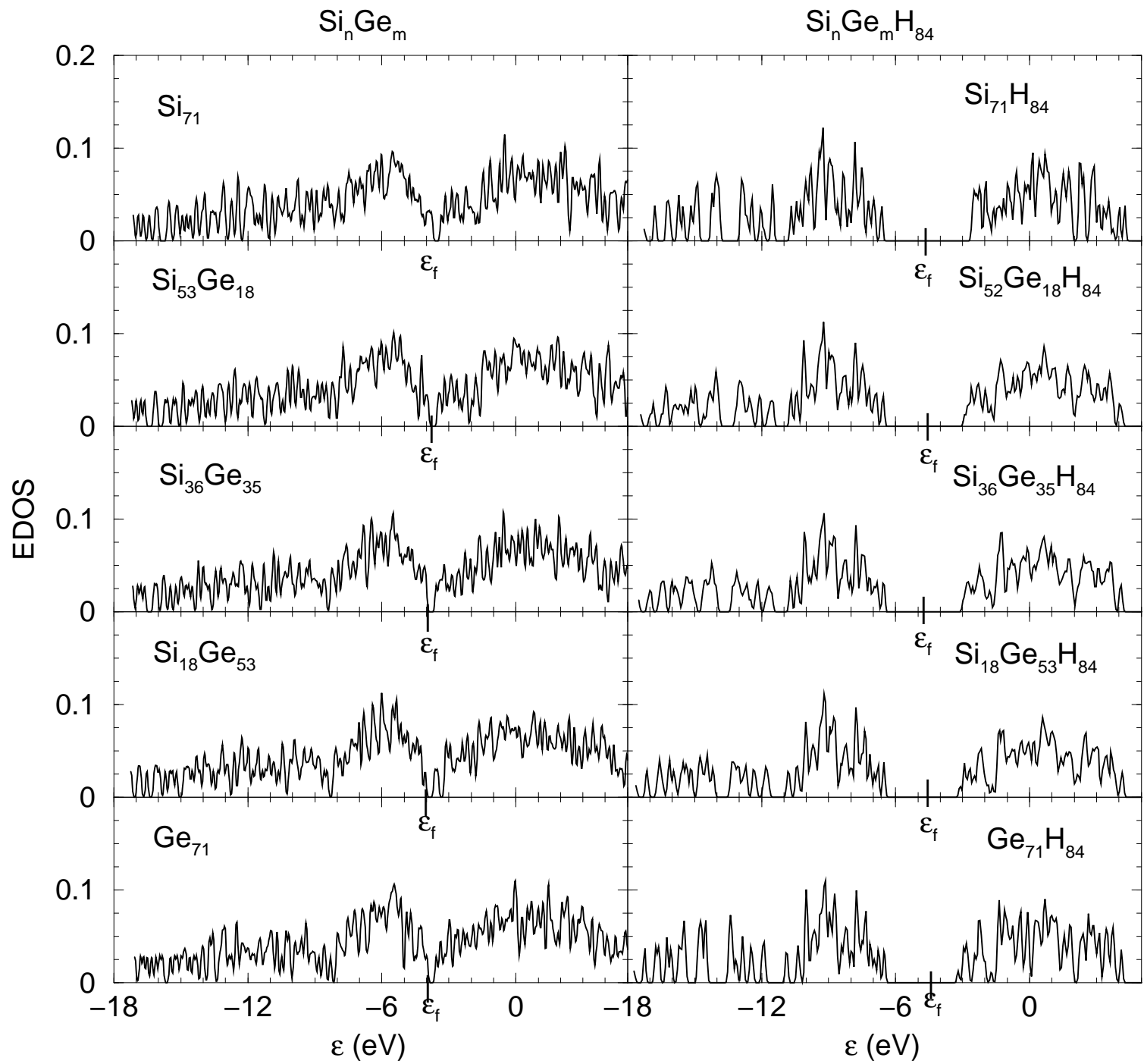
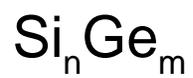
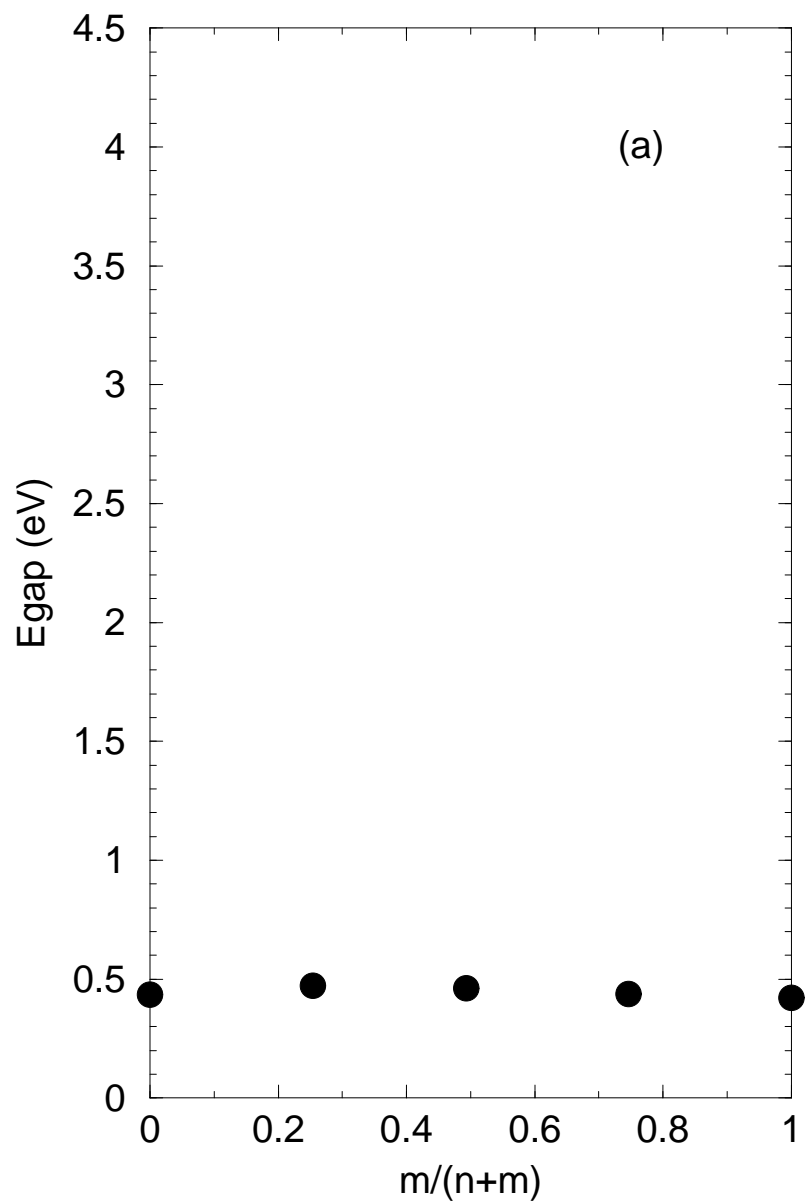
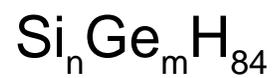
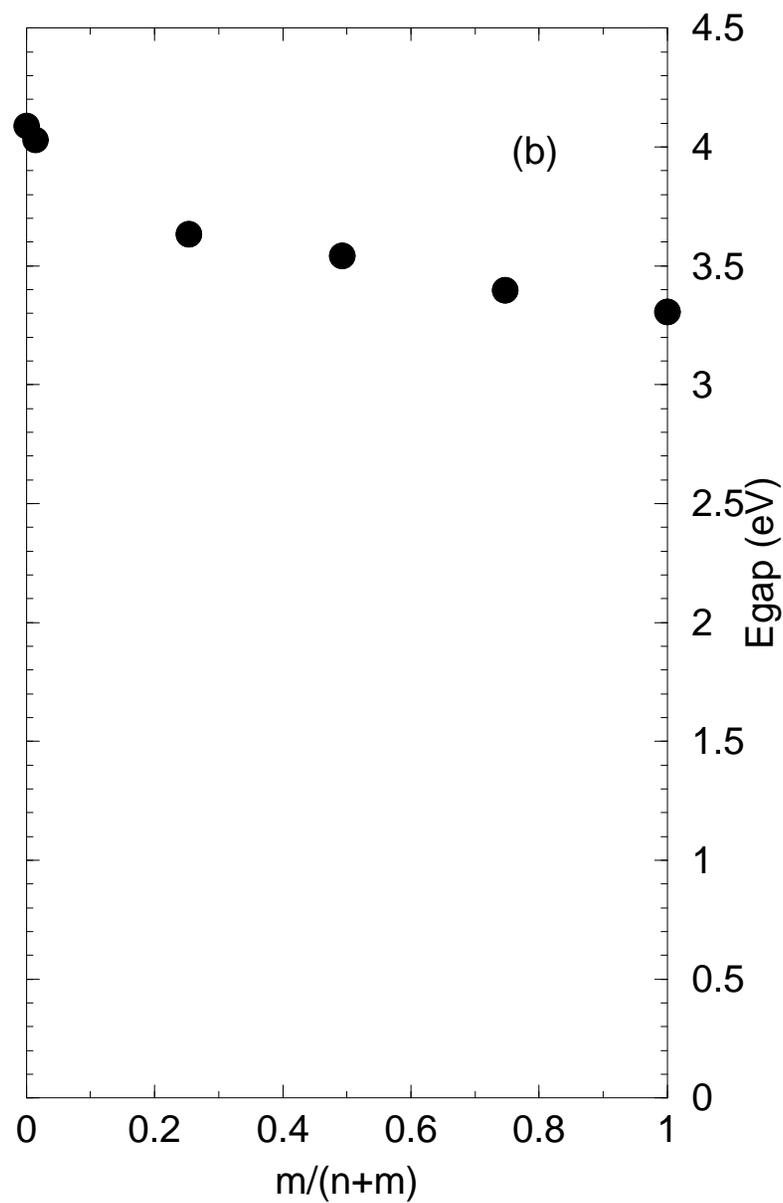
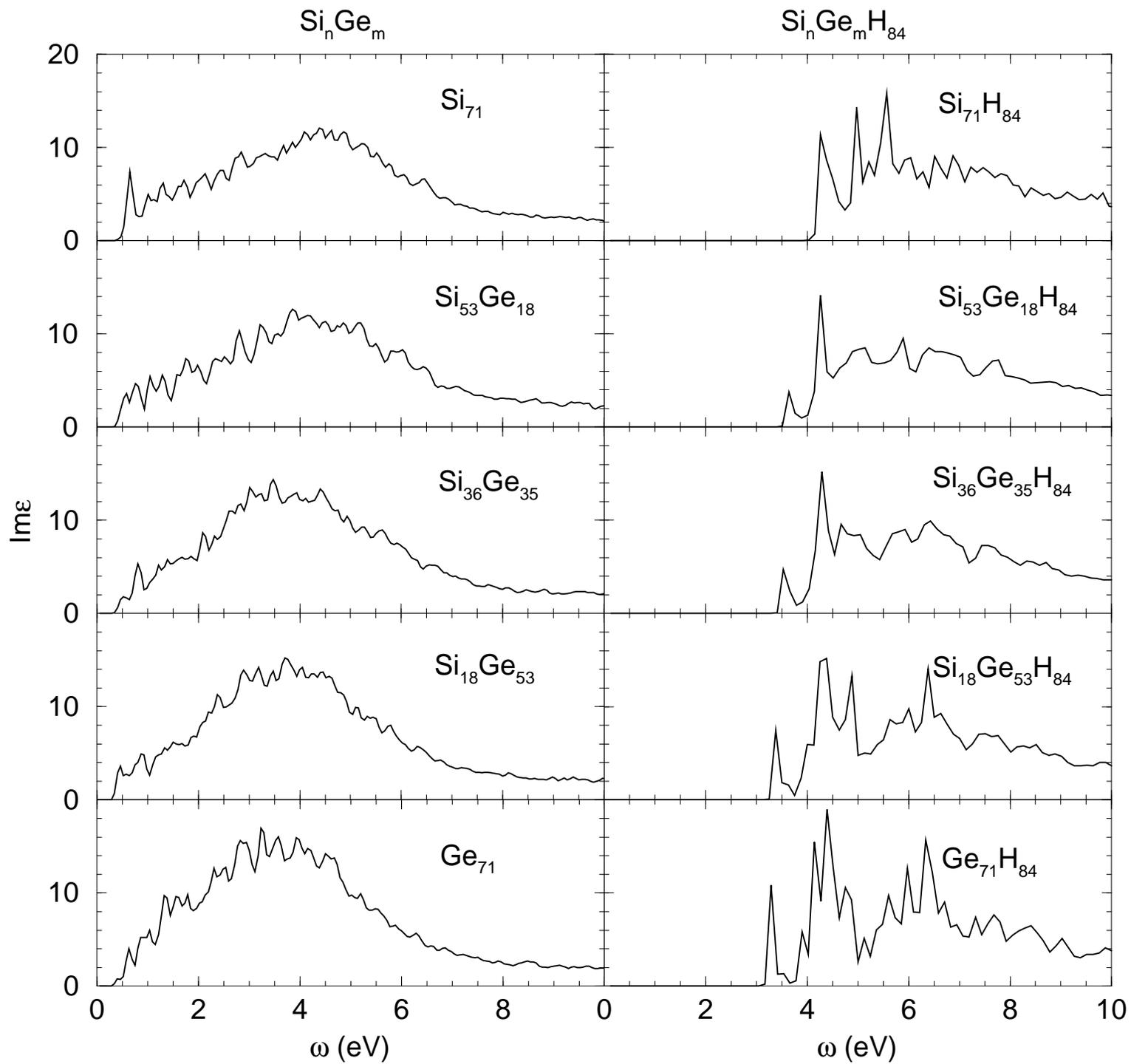
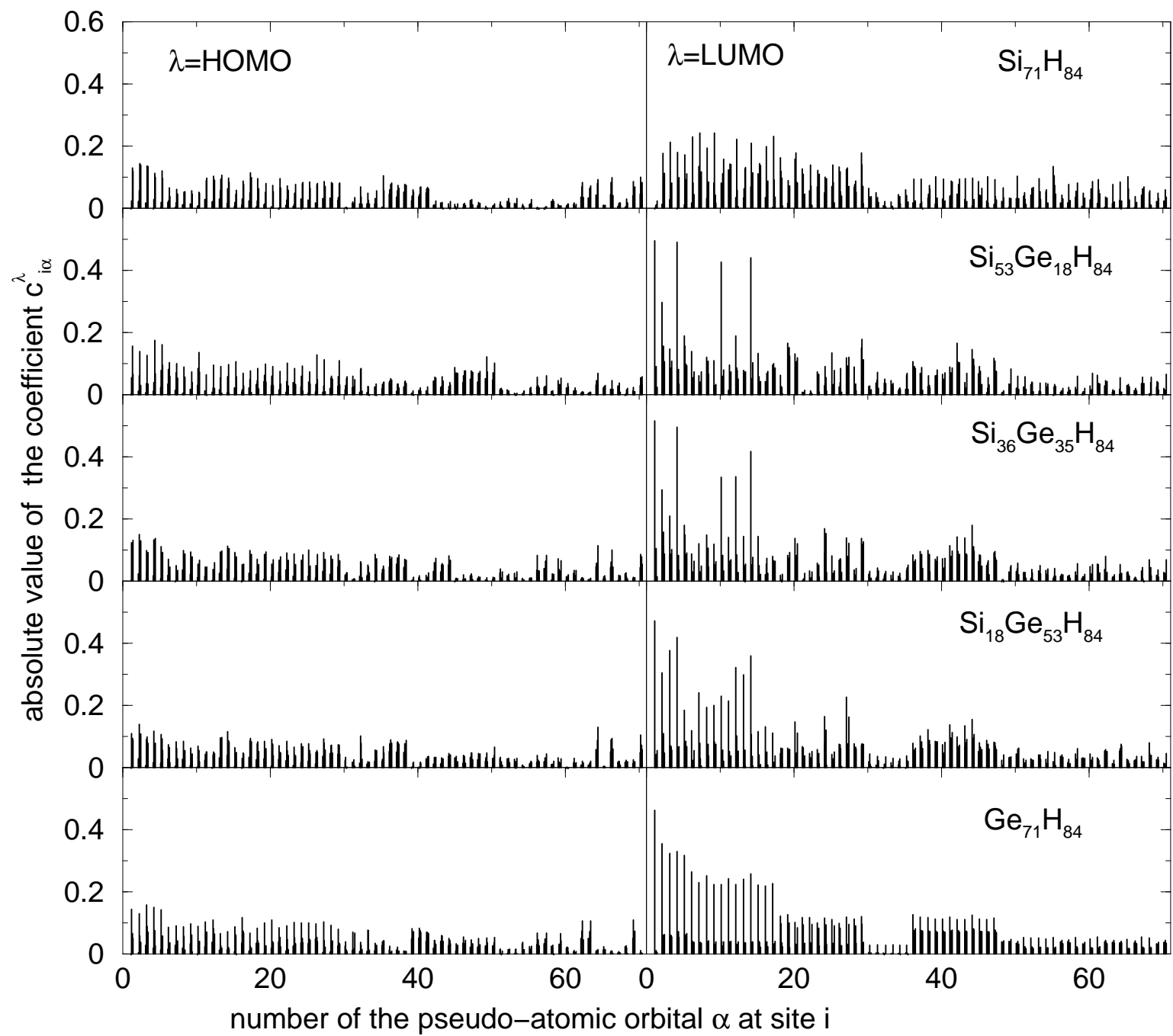
\end{document}